\documentclass{ae100prg}
\newcommand{\reals}{\mathbb{R}}
\usepackage{amsmath}
\usepackage{amssymb}
\usepackage{amsthm}
\newtheorem*{theorem}{Theorem}
\usepackage{times}
\usepackage{graphicx}
\usepackage{wrapfig}
\usepackage{color}
\usepackage{rotating}

\begin{document}
\title{Einstein's ``Prague field-equation''\\
-- another perspective --}

\author{Domenico Giulini$^{1,2}$}

\address{$^1$ Institute for Theoretical Physics\\
Riemann Center for Geometry and Physics\\
Leibniz University of Hannover}
\address{$^2$ Center of Applied Space Technology and Microgravity\\
University of Bremen}

\email{giulini@itp-uni-hannover.de}

\begin{abstract}
I reconsider Einstein's 1912 ``Prague-Theory'' of static 
gravity based on a scalar field obeying a non-linear field 
equation. I point out that this equation follows from 
the self-consistent implementation of the principle that 
\emph{all} energies source the gravitational field according 
to $E=mc^2$. This makes it an interesting toy-model for 
the ``flat-space approach'' to General Relativity (GR), as 
pioneered by Kraichnan and later Feynman. Solutions 
modelling stars show features familiar from GR, e.g., 
Buchdahl-like inequalities. The relation to full GR 
is also discussed. This lends this toy theory also 
some pedagogical significance. This paper is based on a 
talk delivered at the conference \emph{Relativity and Gravitation
100 years after Einstein in Prague}, held in Prague 25.-29. June
2012.

\end{abstract}

\section{Introduction}
Ever since he wrote his large 1907 review of Special Relativity 
\cite{Einstein:1907-b} for the \emph{Jahrbuch der Radioaktivit\"at 
und Elektronik}, Einstein reflected on how to extend the principle of 
relativity to non-inertial motions. His key insight was that such an 
extension is indeed possible, provided gravitational fields are 
included in the description. In fact, the last chapter (V) 
of \cite{Einstein:1907-b}, which comprises four (17-20) out of 
twenty sections, is devoted to this intimate  
relation between acceleration and gravitation. The heuristic 
principle Einstein used was his ``\"Aquivalenzhypothese'' 
(hypothesis of equivalence) or ``\"Aquivalenzprinzip'' 
(principle of equivalence)\footnote{In his Prague papers 
Einstein gradually changed from the first to the second expression.},
which says this: Changing the description of a system from 
an inertial to a non-inertial reference frame is equivalent 
to not changing the frame at all but adding a \emph{special} 
gravitational field. This principle is \emph{heuristic} in the 
sense that it allows to deduce the extension of physical laws, 
the forms of which are assumed to be known in the absence of 
gravitational fields, to the presence of at least those 
special gravitational fields that can be ``created'' by mere 
changes of reference frames. The idea behind this was, 
of course, to postulate that the general features found in 
this fashion remain valid in \emph{all} gravitational fields. 
In the 1907 review Einstein used this strategy to find out about 
the influence gravitational fields have on clocks and general 
electromagnetic processes. What he did not attempt back in 
1907 was to find an appropriate law for the gravitational 
field that could replace the Poisson equation of Newtonian 
gravity. 
This he first attempted in his two ``Prague papers'' from 1912
\cite{Einstein:1912-a}\cite{Einstein:1912-b} for static fields. 
The purpose of my contribution here is to point out that the 
field equation Einstein arrived at in the second of these papers 
is not merely of historical interest. 
   
After 1907 Einstein turned away from gravity research for a 
while, which he resumed in 1911 with a paper~\cite{Einstein:1911-a}, 
also from Prague, in which he used the ``\"Aquivalenzhypothese'' 
to deduce the equality between gravitational and inertial mass, 
the gravitational redshift, and the deflection of light by the 
gravitational field of massive bodies. As is well known, the 
latter resulted in half the amount that was later correctly 
predicted by GR. 

In the next gravity paper~\cite{Einstein:1912-a}, the first 
in 1912, entitled \emph{``Lichtgeschwindigkeit und Statik des 
Gravitationsfeldes''}, Einstein pushed further the consequences 
of his heuristics and began his search for a sufficiently simple 
differential equation for static gravitational fields. 
The strategy was to, first, guess the equation from the form of 
the special fields ``created'' by non inertial reference frames 
and, second, generalise it to those gravitational fields sourced 
by real matter.  Note that the gravitational acceleration was 
to be assumed to be a gradient field (curl free) so that the 
sought-after field equation was for a scalar field, the 
gravitational potential. 

The essential idea in the first 1912 paper is to identify the 
gravitational potential with $c$, the local velocity of light.%
\footnote{Since here we will be more concerned with the 
mathematical form and not so much the actual derivation by 
Einstein, we will ignore the obvious objection that $c$ has 
the wrong physical dimension, namely that of a velocity, 
whereas the a proper gravitational potential should have the 
dimension of a velocity-squared.}
Einstein's heuristics indicated clearly that Special Relativity 
had to be abandoned, in contrast to the attempts by Max Abraham 
(1875-1922), who published a rival theory 
\cite{Abraham:1912-a}\cite{Abraham:1912-b} that was superficially 
based on Poincar\'e invariant equations (but violated Special 
Relativity in abandoning the condition that the four-velocities 
of particles had constant Minkowski square). In passing 
I remark that Einstein's reply \cite{Einstein:1912-c} to Abraham, 
which is his last paper from Prague before his return to Z\"urich,
contains next to his anticipation of the essential physical 
hypotheses on which a future theory of gravity could be based
(here I refer to Ji{\v r}\'{\i} Bi{\v c}{\'a}k's contribution
to this volume), also a concise and very illuminating account 
of the physical meaning and limitation of the special 
principle of relativity, the essence of which was totally 
missed by Abraham. 

Back to Einstein's first 1912 paper, the equation he came up 
with was
\begin{equation}
\label{eq:FirstPragueEquation}
\Delta c=kc\rho\,,
\end{equation}
where $k$ is the ``universal gravitational constant'' and 
$\rho$ is the mass density. The mathematical difference 
between (\ref{eq:FirstPragueEquation}) and the Poisson 
equation in Newtonian gravity is that 
(\ref{eq:FirstPragueEquation}) is homogeneous (even linear) 
in the potential $c$. This means that the source strength of 
a mass density is weighted by the gravitational potential
at its location. This implies a kind of ``red-shift'' for 
the active gravitational mass which in turn results in the 
existence of geometric upper bounds for the latter, as we will 
discuss in detail below. Homogeneity was Einstein's central 
requirement, which he justified from the interpretation of 
the gravitational potential as the local velocity of light, 
which is only determined up to constant rescalings induced 
from rescalings of the timescale.

Already in a footnote referring to equation 
(\ref{eq:FirstPragueEquation}) Einstein points out 
that it cannot be quite correct, as he is to explain in 
detail in a follow-up paper~\cite{Einstein:1912-b}.
This second paper of 1912 is the one I  actually wish 
to focus on in my contribution here. It appeared in the same 
issue of the \emph{Annalen der Physik} as the previous one, 
under the title \emph{``Zur Theorie des statischen Gravitationsfeldes''}
(on the theory of the static gravitational field). In it 
Einstein once more investigates how the gravitational field 
influences electromagnetic and thermodynamic processes 
according to what he now continues to call the `
`\emph{\"Aquivalenzprinzip''}, and derives from it the 
equality of inertial and gravitational mass.%
\footnote{Einstein considers radiation enclosed in a 
container whose walls are ``massless'' (meaning vanishing 
rest-mass) but can support stresses, so as to be able to 
counteract radiation pressure. Einstein keeps repeating 
that equality of both mass types can only be proven if 
the gravitational field does not act on the stressed walls. 
That remark is hard to understand in view of the fact 
that unbalanced stresses add to inertia, as he well knew 
from his own earlier investigations~\cite{Einstein:1907-a}. 
However, as 
explained by Max Laue a year earlier \cite{Laue:1911b},
the gravitational action on the stressed walls is just 
cancelled by that on the stresses of the electromagnetic 
field, for both systems together form a ``complete static 
system'', as Laue calls it. A year later, in the 1913 
``Entwurf'' paper with Marcel Grossmann 
\cite{Einstein-Grossmann:1914}, Einstein again used a 
similar Gedankenexperiment with a massless box containing 
radiation immersed in a gravitational field, by means 
of which he allegedly shows that any Poincar\'e invariant 
scalar theory of gravity must violate energy conservation. 
A modern reader must ask how this can possibly be, in view 
of Noether's theorem applied to time-translation invariance. 
A detailed analysis \cite{Giulini:2008b} shows that this 
energy contains indeed the expected contribution from 
the tension of the walls, which may not be neglected.}

After that he returns to the equation for the static 
gravitational field and considers the gravitational 
force-density $\vec f$, acting on ponderable matter of 
mass density $\rho$, which is given by (Einstein writes $\sigma$ 
instead of our $\rho$) 
\begin{equation}
\label{eq:ForceDensity}
\vec f=-\rho\vec\nabla c\,.
\end{equation} 
Einstein observes that the space integral of $\vec f$ does not 
necessarily vanish on account of (\ref{eq:FirstPragueEquation}),
in violation of the principle that \emph{actio} equals 
\emph{reactio}. Terrible consequences, like self-acceleration, 
have to be envisaged.%
\footnote{\emph{``Anderenfalls w\"urde sich die Gesamtheit der 
in dem betrachteten Raume befindlichen Massen, die wir auf einem 
starren, masselosen Ger\"uste uns befestigt denken wollen, sich 
in Bewegung zu setzen streben.''} (\cite{Einstein:1912-b}, p.\,452)} 
He then comes up with the following non-linear but still homogeneous modification 
of (\ref{eq:FirstPragueEquation}) (again Einstein writes $\sigma$
instead of $\rho$): 
\begin{equation}
\label{eq:SecondPragueEquation}
\Delta c=k
\left\{
c\rho+\frac{1}{2k}\frac{\vec\nabla c\cdot\vec\nabla c}{c}
\right\}\,.
\end{equation}

In the rest of this paper we will show how to arrive at 
this equation from a different direction and discuss 
some of its interesting properties as well as its 
relation to the description of static gravitational fields 
in GR. 

\section{A self-consistent modification of Newtonian Gravity}
\label{sec:SelfConsistentMod}
The following considerations are based on \cite{Giulini:1997c}.
We start from ordinary Newtonian gravity, where the gravitational 
field is described by a scalar function $\varphi$ whose physical 
dimension is that of a velocity-squared. It obeys 
\begin{equation}
\label{eq:NewtonianFieldEq}
\Delta\varphi=4\pi G\,\rho\,.
\end{equation}
The force per unit volume that the gravitational field 
exerts onto a distribution of matter with density $\rho$ is  
\begin{equation}
\label{eq:NewtonianForceDensity}
\vec f=-\rho\vec\nabla\varphi\,.
\end{equation}

This we apply to the force that the gravitational field 
exerts onto its own source during a real-time process of 
redistribution. This we envisage as actively transporting each 
mass element along the flow line of a vector field 
$\vec\xi$. To first order, the 
change $\delta\rho$ that $\rho$ suffers in time $\delta t$ 
is given by 
\begin{equation}
\label{eq:DensityChange}
\delta\rho
=\frac{-L_{\delta \vec\xi}\bigl(\rho d^3x\bigr)}{d^3x}   
=-\vec\nabla\cdot (\delta\vec\xi\,\rho)\,,
\end{equation}
where $\delta\vec\xi=\delta t\,\vec\xi$ and $L_{\delta \vec\xi}$
is the Lie derivative with respect to $\delta\vec\xi$. 
We assume the  support $\text{supp}(\rho)=:B\subset\reals^3$ 
to be compact. In general, this redistribution costs energy. 
The work we have to invest for redistribution is, to first order,  
just given by
\begin{equation}
\label{eq:InvestedWorkGeneral}
\delta A
=-\int_{\reals^3}\delta\vec\xi\cdot\vec f
=-\int_B\varphi\vec\nabla\cdot(\delta\vec\xi\,\rho)
=\int_B\varphi\,\delta\rho\,,
\end{equation}
where we used (\ref{eq:DensityChange}) in the last step
and where we did not write out the Lebesgue measure $d^3x$
to which all integrals refer. Note that in order to obtain 
(\ref{eq:InvestedWorkGeneral}) we did not make use the field 
equation. Equation (\ref{eq:InvestedWorkGeneral}) is generally 
valid whenever the force-density relates to the potential 
and the mass density as in (\ref{eq:NewtonianForceDensity}). 

Now we make use of the field equation (\ref{eq:NewtonianFieldEq}).
We assume the redistribution-process to be adiabatic, that is, 
we assume the instantaneous validity of the field equation 
at each point in time throughout the process. This implies  
\begin{equation}
\label{eq:AdiabaticAssumption}
\Delta\delta\varphi=4\pi G\,\delta\rho\,.
\end{equation}
Hence, using (\ref{eq:InvestedWorkGeneral}), 
the work invested in the process of redistribution 
is (to first order) 
\begin{equation}
\label{eq:InvestedWorkNewton}
\delta A=\int_B\varphi\,\delta\rho=
\delta\left\{
-\frac{1}{8\pi G}\int_{\reals^3}(\vec\nabla\varphi)^2
\right\} \,.
\end{equation}    
If the infinitely dispersed state of matter is assigned the 
energy-value zero, then the expression in curly brackets is the 
total work invested in bringing the infinitely dispersed state 
to that described by the distribution $\rho$. This work must 
be stored somewhere as energy. Like in electro-statics and 
-dynamics, we take a further logical step and assume this 
energy to be spatially distributed in the field according 
to the integrand. This leads to the following expression 
for the energy density of the static gravitational field   
\begin{equation}
\label{eq:NewtonianEnergyDensity}
\varepsilon=-\frac{1}{8\pi G}(\vec\nabla\varphi)^2\,.
\end{equation}    
All this is familiar from Newtonian gravity. But now we go 
beyond Newtonian gravity and require the validity of the 
following 

\medskip\noindent 
\textbf{Principle.}
\emph{All energies, including that of the gravitational field 
itself, shall gravitate according to $E=mc^2$.}
\medskip

\noindent
This principle implies that if we invest an amount of work 
$\delta A$ to a system its (active) gravitational mass will 
increase by $\delta A/c^2$. 

Now, the (active) gravitational mass $M_g$  is defined by 
the flux of the  gravitational field to spatial infinity (i.e. 
through spatial spheres as their radii tend to infinity): 
\begin{equation}
\label{eq:DefGravMass}
M_g
=\frac{1}{4\pi G}\int_{S^2_\infty}\vec n\cdot\vec\nabla\varphi
=\frac{1}{4\pi G}\int_{\reals^3}\Delta\varphi\,.
\end{equation}
Hence, making use of the generally valid equation 
(\ref{eq:InvestedWorkGeneral}), the principle that 
$\delta A=M_gc^2$ takes the form 
\begin{equation}
\label{eq:ThePrinciple}
\int_B\varphi\,\delta\rho
=\frac{c^2}{4\pi G}\int_{\reals^3}\Delta\delta\varphi\,.
\end{equation}
This functional equation relates $\varphi$ and $\rho$, 
over and above the restriction imposed on their relation 
by the field equation. However, the latter may - and 
generally will - be inconsistent with this additional 
equation. For example, the Newtonian field equation 
(\ref{eq:NewtonianFieldEq}) is easily seen to 
manifestly violate (\ref{eq:ThePrinciple}), for the 
right-hand side then becomes just the integral over 
$c^2\delta\rho$, which always vanishes on account of 
(\ref{eq:DensityChange}) (or the obvious remark that 
the redistribution clearly does not change the total 
mass), whereas the left 
hand side will generally be non-zero. 
The task must therefore be to find field equation(s) 
consistent with (\ref{eq:ThePrinciple}). Our main 
result in that direction is that the unique 
generalisation of (\ref{eq:NewtonianFieldEq}) which satisfies
(\ref{eq:ThePrinciple}) is just (\ref{eq:SecondPragueEquation}),
i.e. the field equation from Einstein's second 1912 paper.  

Let us see how this comes about. A first guess for a consistent 
modification of (\ref{eq:NewtonianFieldEq}) is to simply add 
$\varepsilon/c^2$ to the source $\rho$:
\begin{equation}
\label{eq:ImprovedNG-FirstIteration}
\Delta\varphi=
4\pi G\left(\rho-\frac{1}{8\pi Gc^2}
\bigl(\nabla\varphi\bigr)^2\right)\,.
\end{equation}
But this cannot be the final answer because this change of the 
field equation also brings about a change in the expression 
for the self-energy of the gravitational field. That is, the term
in the bracket on the right-hand side is not the total energy 
according to \emph{this} equation, but according to the original 
equation (\ref{eq:NewtonianFieldEq}). 
In other words: equation (\ref{eq:ImprovedNG-FirstIteration})
still lacks \emph{self-}consistency. This can be corrected 
for by iterating this procedure, i.e., determining the field's 
energy density according to (\ref{eq:ImprovedNG-FirstIteration}) 
and correcting the right-hand side of 
(\ref{eq:ImprovedNG-FirstIteration}) accordingly. Again we have 
changed the equation, and this goes on ad infinitum. But the 
procedure converges to a unique field equation, similarly to the
convergence of the ``Noether-procedure''%
\footnote{Pioneered by Robert Kraichnan in his 1947 MIT Bachelor 
thesis ``Quantum Theory of the Linear Gravitational Field''.}
that leads from the Poincar\'e invariant Pauli-Fierz theory of 
spin-2 mass-0 fields in flat Minkowski space to GR
\cite{Kreichnan:1955}\cite{Feynman}\cite{Deser:1970}. 

In our toy model the convergence of this procedure is not  
difficult to see.  We start from the definition 
(\ref{eq:DefGravMass}) and calculate its variation 
$\delta M_g$ assuming the validity of (\ref{eq:ImprovedNG-FirstIteration}).
From what we said above we know already that this not yet going to 
satisfy (\ref{eq:ThePrinciple}). But we will see that from 
this calculation we can read off the right redefinitions. 

We start by varying (\ref{eq:DefGravMass}):
\begin{equation}
\label{eq:ZerothIteration}
\delta M_g=\frac{1}{4\pi G}\int_{\reals^3}\Delta\delta\varphi\,.
\end{equation}
We replace $\Delta\delta\varphi$ with the variation 
of the right-hand side of (\ref{eq:ImprovedNG-FirstIteration}). 
Partial integration of the non-liner part gives us 
a surface term whose integrand is 
$\propto \varphi\vec\nabla\delta\varphi=O(r^{-3})$ and 
hence vanishes. The remaining equation is 
\begin{equation}
\label{eq:FirstIteration}
\delta M_g=\int_B\delta\rho+\frac{1}{4\pi G}\int_{\reals^3} 
\left(\frac{\varphi}{c^2}\right)\Delta\delta\varphi\,.
\end{equation}
Playing the same trick (of replacing $\Delta\delta\varphi$ 
with the variation of the right-hand side of 
(\ref{eq:ImprovedNG-FirstIteration}) and partial 
integration, so as to collect all derivatives on 
$\delta\varphi$) again and again, we arrive after 
$N$ steps at 
\begin{equation}
\label{eq:nthIteration}
\delta M_g
=\int_B\ \sum_{n=0}^{N-1}\frac{1}{n!}\left(\frac{\varphi}{c^2}\right)^n\delta\rho
+\frac{1}{N!c^{2N}}\frac{1}{4\pi G}\int_{\reals^3}\varphi^N\delta(\Delta\varphi)\,.
\end{equation}
As $\varphi$ is bounded for a regular matter distribution, 
and the spatial integral over $\delta\Delta\varphi$ is just 
$4\pi G\delta M_g$, the last term tends to zero for 
$N\rightarrow\infty$. Hence 
\begin{equation}
\label{eq:LastStep}
\delta M_g=\int_B\delta\rho\,\exp(\varphi/c^2)\,.
\end{equation}
This \emph{is} of the desired form (\ref{eq:ThePrinciple})
required by the principle, provided we redefine the 
gravitational potential to be $\Phi$ rather than $\varphi$,
where 
\begin{equation}
\label{eq:NewPotential}
\Phi:=c^2\,\exp(\varphi/c^2)\,.
\end{equation}
Saying that $\Phi$ rather than $\varphi$ is the right 
gravitational potential means that the force density 
is not given by (\ref{eq:NewtonianForceDensity}), 
but rather by
\begin{equation}
\label{eq:NewForceDensity}
\vec f=-\rho\vec\nabla\Phi\,.
\end{equation}
As we have made use of equation 
(\ref{eq:ImprovedNG-FirstIteration}) in order to 
derive (\ref{eq:LastStep}), we must make sure to 
keep \emph{that} equation, just re-expressed in 
terms of $\Phi$. This leads to 
\begin{equation}
\label{eq:SelfConsistentEq}
\Delta\Phi
=\frac{4\pi G}{c^2}
\left[
\rho\Phi+\frac{c^2}{8\pi G}
\frac{(\vec\nabla\Phi)^2}{\Phi}
\right]\,,
\end{equation}
which is precisely Einsteins improved ``Prague equation'' 
(\ref{eq:SecondPragueEquation}) with $k=4\pi\,G/c^2$.
Note from (\ref{eq:NewPotential}) that the asymptotic 
condition $\varphi(r\rightarrow\infty)\rightarrow 0$ 
translates to $\Phi(r\rightarrow\infty)\rightarrow c^2$. 
Note also that for $r\rightarrow\infty$ the $1/r^2$-parts 
of $\vec\nabla\varphi$ and $\vec\nabla\Phi$ coincide,
so that in the expressions (\ref{eq:DefGravMass}) for 
$M_g$ we may just replace $\varphi$ with $\Phi$: 
\begin{equation}
\label{eq:DefGravMassNew}
M_g
=\frac{1}{4\pi G}\int_{S^2_\infty}\vec n\cdot\vec\nabla\Phi
=\frac{1}{4\pi G}\int_{\reals^3}\Delta\Phi\,.
\end{equation}
The principle now takes the form (\ref{eq:ThePrinciple})
with $\varphi$ replaced by $\Phi$. It is straightforward 
to show by direct calculation that (\ref{eq:ThePrinciple}) 
is indeed a consequence of (\ref{eq:SelfConsistentEq}), as it must 
be. It also follows from  (\ref{eq:SelfConsistentEq}) that 
the force density (\ref{eq:NewForceDensity}) is the divergence 
of a symmetric tensor:
\begin{subequations}
\label{eq:DivSymmTensor}
\begin{equation}
\label{eq:DivSymmTensor-a}
f_a=-\nabla^bt_{ab}\,,
\end{equation}
where
\begin{equation}
\label{eq:DivSymmTensor-b}
t_{ab}=\frac{1}{4\pi Gc^2}\left\{\frac{1}{\Phi}
\left[
\nabla_a\Phi\nabla_b\Phi-\tfrac{1}{2}\delta_{ab}(\vec\nabla\Phi)^2
\right]
\right\}\,.
\end{equation}
\end{subequations}
This implies the validity of the principle that actio equals reactio 
that Einstein demanded. \emph{This} was Einstein's rationale 
for letting (\ref{eq:SecondPragueEquation}) replace 
(\ref{eq:FirstPragueEquation}).  

Finally we mention that (\ref{eq:SelfConsistentEq}) may be 
linearised if written in terms of the square-root of $\Phi$:
\begin{equation}
\label{eq:LinearisedSelfConsistentEq-1}
\Psi:=\sqrt{\frac{\Phi}{c^2}}\,.
\end{equation} 
One gets
\begin{equation}
\label{eq:SelfConsistentEqLinear}
\Delta\Psi
=\frac{2\pi G}{c^2}\,\rho\,\Psi\,.
\end{equation}
This helps in finding explicit solutions to 
(\ref{eq:SelfConsistentEq}). Note that $\Psi$ is
dimensionless.

\section{Spherically symmetric solutions}
In this section we discuss some properties of spherically symmetric 
solutions to (\ref{eq:SelfConsistentEqLinear}) for spherically 
symmetric mass distributions $\rho$ of compact support. In the following 
we will simply refer to the object described by such a mass distribution 
as ``star''. 

In terms of $\chi(r):=r\Psi(r)$ equation (\ref{eq:SelfConsistentEqLinear}) 
is equivalent to
\begin{equation}
\label{eq:SelfConsistentEqLinear-Chi}
\chi''=\frac{2\pi G}{c^2}\,\rho\,\chi\,.
\end{equation}
The support of $\rho$ is a closed ball of radius $R$, called 
the star's radius. For $r<R$ we shall assume $\rho(r)\geq 0$
(weak energy condition). We seek solutions which correspond to 
everywhere positive and regular $\Psi$ and hence everywhere 
positive and regular $\Phi$. In particular $\Phi(r=0)$ and 
$\Psi(r=0)$ must be finite. For $r>R$ equation 
(\ref{eq:SelfConsistentEqLinear-Chi}) implies $\chi''=0$,
the solution to which is   
\begin{equation}
\label{eq:PsiLargeR}
\chi_+(r)=r\Psi_+(r)=r-R_g\,,\quad\text{for}\ r>R\,,
\end{equation}
where $R_g$ denotes the gravitational radius
\begin{equation}
\label{eq:DefGravRadius}
R_g:=\frac{GM_g}{2c²}\,.
\end{equation}
$R_g$ comes in because of (\ref{eq:DefGravMassNew}),
which fixes one of the two integration constants, the 
other being fixed by $\Psi(\infty)=1$.  

Let $\chi_-$ denote the solution in the interior of the star.
Continuity and differentiability at $r=R$ gives $\chi_-(R)=R-R_g$ 
and ${\chi'}_-(R)=1$. We observe that $\chi_-(R)\geq 0$. For suppose
$\chi_-(R)<0$, then (\ref{eq:SelfConsistentEqLinear-Chi}) and 
the weak energy condition imply $\chi''(R)\leq 0$. But this 
implies that for $r\in[0,R]$ the curve $r\mapsto\chi_-(r)$ lies 
below the straight line $r\mapsto r-R_g$ and assumes a value 
less than $-R_g$ at $r=0$, in contradiction to the finiteness of 
$\Psi(r=0)$ which implies $\chi_-(r=0)=0$. Hence we have  

\begin{theorem}
The gravitational radius of a spherically symmetric star 
is universally bound by its (geometric) radius, $R_g\leq R$.
Equivalently expressed in terms of $M_g$ we may say that 
the gravitational mass is universally bound above by 
\begin{equation}
\label{eq:UnivMassUpperBound}
M_g<\frac{2c^2R}{G}\,.
\end{equation} 
\end{theorem}

This may be seen in analogy to Buchdahl's inequality in 
GR \cite{Buchdahl:1959}, which, using the 
isotropic (rather than Schwarzschild) radial coordinate, 
would differ from (\ref{eq:UnivMassUpperBound}) only by 
an additional factor of $8/9$ on the right-hand side. 
The Buchdahl bound is optimal, being saturated by the 
interior Schwarzschild solution for a homogeneous star.

So let us here, too, specialise to a homogeneous
star, 
\begin{equation}
\label{eq:HomogeneousStar-Density}
\rho(r)=
\begin{cases}
\frac{3M_b}{4\pi R^3} &\text{for}\ r\leq R\\
0&\text{for}\ r> R\,,
\end{cases}
\end{equation}
where $M_b$ is called the bare mass (integral over $\rho$). 
It is convenient to introduce the radii corresponding to bare and 
gravitational masses, as well as their ratio to the star's 
radius $R$:
\begin{subequations}
\label{eq:DefRadii}
\begin{alignat}{4}
\label{eq:DefRadii-a}
&R_b&&:=\frac{GM_b}{2c^2}\,,\quad &&x &&:=\frac{R_b}{R}\,,\\
&R_g&&:=\frac{GM_g}{2c^2}\,,\quad &&y &&:=\frac{R_g}{R}\,.
\end{alignat}
\end{subequations}
We also introduce the inverse length
\begin{equation}
\label{eq:DefOmega}
\omega:=\frac{1}{R}\cdot\sqrt{\frac{3R_b}{R}}\,,
\end{equation}
so that for (\ref{eq:SelfConsistentEqLinear-Chi}) just reads 
$\chi''=\omega^2\chi$. From this the interior solution is 
easily obtained. If written in terms of $\Psi$ it reads  
\begin{equation}
\label{eq:InteriorSolutionPsi}
\Psi_-(r)=\frac{1}{\cosh(\omega R)}\frac{\sinh(\omega r)}{\omega r}\,,
\quad\text{for}\ r<R\,.
\end{equation}
As a result of the matching to the exterior solution given 
in (\ref{eq:PsiLargeR}), $R_g$ is determined by $R$ and 
$\omega$, i.e. $R$ and $R_b$. In terms of $x$ and $y$  this relation takes 
the simple form 
\begin{equation}
\label{eq:GravMassExpression}
y=1-\frac{\tanh\bigl(\sqrt{3x}\bigr)}{\sqrt{3x}}\,,
\end{equation}
which convex-monotonically maps $[0,\infty)$ onto $[0,1)$. 
The fact that $y<1$ for all $x$ is just the statement 
of the Theorem applied to the homogeneous case. 

If $x=R_b/R\ll 1$ we have $y=x-\tfrac{6}{5}x^2+\cdots$,
which for $E_{\rm total}:=M_gc^2$ reads 
\begin{equation}
\label{eq:GravMassApprox}
E_{\rm total}=M_bc^2\left(1-\tfrac{3}{5}x+O(x^2)\right)\,.
\end{equation}
We note that $-3M_bc^2x/5=-\tfrac{3}{5}GM_b^2/R$ is just the Newtonian 
binding energy of a homogeneous star. In view of our Principle 
it makes good sense that to first order just this amount is 
subtracted from the bare mass in order to obtain the active gravitational mass. In Newtonian gravity this negative amount is just identified 
with the field's self-energy, but here the interpretation is different: 
The two terms that act as source for the gravitational field in (\ref{eq:SelfConsistentEq})
are the matter part, which is proportional to $\rho$ but diminished 
by $\Phi$, and the field's own part, which is proportional to 
$(\vec\nabla\Phi)^2/\Phi$ and positive definite! Their contributions 
are, respectively, 
\begin{alignat}{2}
&E_{\rm matter}&&=\int_B\rho\Phi=M_bc^2\left(1-\tfrac{6}{5}x+O(x^2)\right)\,,\\
&E_{\rm field}&&=\frac{c^2}{8\pi G}\int_{\reals^3}\frac{(\vec\nabla\Phi)^2}{\Phi}=
M_bc^2\left(\tfrac{3}{5}x+O(x^2)\right)\,.
\end{alignat}
Hence even though the total energy is decreased due to binding,
the gravitational field's self energy \emph{increases} by the 
same amount. Twice that amount is gained from the fact that the 
matter-energy is ``red-shifted'' by being multiplied with $\Phi$,
so energy is conserved (of course). 

Two more consequences, which are related, are noteworthy: 
\begin{itemize}
\item
Unlike in Newtonian theory, objects with non-zero gravitational 
mass cannot be modelled by point sources. In the spherically 
symmetric case this is an immediate consequence  of 
(\ref{eq:UnivMassUpperBound}), which implies 
$M_g\rightarrow 0$ for $R\rightarrow 0$. Hence there are no 
$\delta$-like masses. 
\item
Unlike in Newtonian gravity, unlimited compression of matter 
does not lead to unlimited energy release. Consider a sequence 
of homogeneous (just for simplicity) stars of fixed bare mass 
$M_b$ and variable radius $R$, then the gravitational mass $M_g$ 
as function of $x=R_b/R$ is given by 
\begin{equation}
\label{eq:GravMassFunctR}
M_g(x)=M_b\cdot
\left\{\frac{1}{x}\cdot
\left(1-\frac{\tanh\left(\sqrt{3x}\right)}{\sqrt{3x}}\right)\right\}\,.
\end{equation}
The function in curly brackets\footnote{Its Taylor expansion at $x=0$ 
is $1-6x/5+51x^2/35+\cdots$.} is a strictly monotonically decreasing 
function $[0,\infty]\mapsto[1,0]$. This shows that for infinitely 
dispersed matter, where $R\rightarrow\infty$ and hence $x\rightarrow 0$, 
we have $M_g(x=0)=M_b$, as expected, and that for infinite 
compression $M_g(x\rightarrow\infty)=0$. As the gained energy at 
stage $x$ is $(M_b-M_g(x))c^2$, we can at most gain $M_bc^2$.
\end{itemize}

\section{Relation to General Relativity}
Finally I wish to briefly comment on the relation of 
equation (\ref{eq:SecondPragueEquation}) or 
(\ref{eq:SelfConsistentEq}) to GR. 
Since Einstein's 1912 theory was only meant to be valid
for static situations, I will restrict attention to 
static spacetimes $(M,g)$. Hence I assume the existence 
of a timelike and hypersurface orthogonal Killing field 
$K$. My signature convention shall be ``mostly plus'', 
i.e. $(-,+,+,+)$. 

We choose adapted coordinates $(t,x^a)$, $a=1,2,3$, where the 
level sets of $t$ are the integral manifolds of the foliation
defined by $K$ and $K=\partial/\partial(ct)$. We can then write 
the metric in a form in which the coefficients do not depend 
on $t$ (called ``time'') ,   
\begin{equation}
\label{eq:StaticMetric}
g=-\Psi^2(x)\,c^2\,dt\otimes dt +{\hat g}_{ab}(x)\,dx^a\otimes dx^b\,.
\end{equation}
Clearly $c^2\Psi^2=-g(K,K)$. From now on, all symbols 
with hats on refer to the spatial geometry, like the 
spatial metric $\hat g$.   

The $t$-component of the geodesic equation is equivalent to  
$\Psi^2\dot t=\text{const}$, where an overdot refers to the 
derivative with respect to an affine parameter. This equation 
allows us to eliminate the affine parameter in favour of $t$ 
in the spatial components of the geodesic equation. If we set%
\footnote{This differs by a factor of 2 from 
(\ref{eq:LinearisedSelfConsistentEq-1}) which we need and 
to which we return below.} 

\begin{equation}
\label{eq:PhiPsiRelation}
\Psi=\sqrt{\frac{2\Phi}{c^2}}
\end{equation}
they read 
\begin{equation}
\label{eq:GeodesicEqSpatial}
\frac{d^2 x^a}{dt^2}
+{\hat \Gamma}^a_{bc}\frac{dx^b}{dt}\frac{dx^c}{dt}
=-\Phi_{,b}{\hat g}^{ab}+\Phi_{,b}
\left[\frac{1}{\Phi}\frac{dx^a}{dt}\frac{dx^b}{dt}\right]\,,
\end{equation}
where the ${\hat\Gamma}^a_{bc}$ are the Christoffel 
coefficients for $\hat g$, and $\Phi_{,a}=\partial_a\Phi$. 
This should be compared with (\ref{eq:NewForceDensity}) together with Newton's second 
law, which give $d^2\vec x/dt^2=-\vec\nabla\Phi$.
As we did not attempt to include special relativistic 
effects in connection with high velocities, we should 
consistently neglect terms $v^2/c^2$ in 
(\ref{eq:GeodesicEqSpatial}). This results in dropping 
the rightmost term. The rest has the pseudo-Newtonian form
in arbitrary (not just inertial) spatial coordinates.
A non-zero spatial curvature would, of course, be a new 
feature not yet considered. 

The curvature and Ricci tensors for the metric 
(\ref{eq:StaticMetric}) are readily computed, most 
easily by using Cartan's structure equations: 
\begin{equation}
\label{eq:RicciTensor}
\text{Ric}(n,n)=\Psi^{-1}\,\hat\Delta\Psi\,,
\quad
R_{ab}={\hat R}_{ab}-\Psi^{-1}\,\hat\nabla_a\hat\nabla_b\Psi\,.
\end{equation}
Here $n=\Psi^{-1}\partial/c\partial t$ is the unit timelike 
vector characterising the static reference frame, $\hat\nabla$ 
is the Levi-Civita covariant derivative with respect to $\hat g$, 
and $\hat\Delta$ is the corresponding Laplacian.  

Using this in Einstein's equations 
\begin{equation}
\label{eq:EinsteinsEquationsGeneral}
R_{\mu\nu}=\frac{8\pi G}{c^4}
\Bigl(T_{\mu\nu}-\tfrac{1}{2}g_{\mu\nu}T^\lambda_\lambda\Bigr)
\end{equation}
for pressureless (we neglect the pressure since it enters multiplied 
with $c^{-2}$) dust at rest and of mass-density $\rho$ in the static 
frame, i.e.
\begin{equation}
\label{eq:PressurlessDust}
T_{\mu\nu}=\rho c^2 n_\mu n_\nu\,,
\end{equation}
we get 
\begin{subequations}
\label{eq:EinsteinsEquationsHere}
\begin{alignat}{2}
\label{eq:EinsteinsEquationsHere-a}
\hat\Delta\Psi
&=\frac{4\pi G}{c^2}\rho\Psi\qquad
&&\text{time component}\,,\\
\label{eq:EinsteinsEquationsHere-b}
\hat\nabla_{a}\hat\nabla_{b}\Psi
&=\hat R_{ab}\Psi\qquad
&&\text{space components}\,.
\end{alignat}
\end{subequations}

We note that, apart from the space curvature, 
(\ref{eq:EinsteinsEquationsHere-a}) is almost---but not 
quite---identical to (\ref{eq:SelfConsistentEqLinear}). They differ 
by a factor of 2! Rewriting (\ref{eq:EinsteinsEquationsHere-a}) 
in terms of $\Phi$ according to (\ref{eq:PhiPsiRelation}), 
we get 
\begin{equation}
\label{eq:SelfConsistentEqGR}
\hat\Delta\Phi
=\frac{8\pi G}{c^2}
\left[
\rho\Phi+\frac{c^2}{16\pi G}
\frac{{\hat g}^{ab}\hat\nabla_a\Phi\hat\nabla_b\Phi}{\Phi}
\right]\,.
\end{equation}
This differs from (\ref{eq:SelfConsistentEq}) by the same factor 
of $2$ (i.e., $G\rightarrow 2G$). Note that we cannot simply remove 
this factor by rescaling 
$\Psi$ and $\Phi$, as the equations are homogeneous in these 
fields. Note also that the overall scale of $\Phi$ is fixed by 
(\ref{eq:GeodesicEqSpatial}): It is the gradient of 
$\Phi$, and not a multiple thereof, which gives the 
acceleration. But then there is another factor of 2
in difference to our earlier discussion: If the 
metric (\ref{eq:StaticMetric}) is to approach the 
Minkowski metric far away from the source, then $\Psi$ 
should tend to one and hence $\Phi$ should 
asymptotically approach $c^2/2$ according to 
(\ref{eq:PhiPsiRelation}). In (\ref{eq:SelfConsistentEq}),
however, $\Phi$ should asymptotically approach $c^2$, 
i.e. twice that value. This additional factor of 2 
ensures that both theories have the same Newtonian limit. 
Indeed, if we expand the gravitational potential $\Phi$ 
of an isolated object in a power series in $G$, this 
implies that the linear terms of both theories coincide.
However, the quadratic terms in GR are twice as large as 
in our previous theory based on (\ref{eq:NewForceDensity}) 
and (\ref{eq:SelfConsistentEq}). This is not quite unexpected 
if we take into account that in GR we also have the space 
curvature that will modify the fields and geodesics in post 
Newtonian approximations. We note that the spatial Einstein 
equations (\ref{eq:EinsteinsEquationsHere-b}) prevent space 
from being flat. For example, taking  their trace and using 
(\ref{eq:EinsteinsEquationsHere-a}) shows that the scalar 
curvature of space is, in fact,  proportional to the mass density. 

Finally we show that the total gravitational mass in GR is just 
given by the same formula (\ref{eq:DefGravMassNew}), where $\Phi$
is now that used here in the GR context . To see this we recall 
that for spatially asymptotically flat spacetimes the overall mass 
(measured at spatial infinity) is given by the ADM-mass. Moreover, 
for spatially asymptotically flat spacetimes which are stationary 
and satisfy Einstein's equations with sources of spatially compact 
support, the ADM mass is given by the Komar integral (this is, e.g., 
proven in Theorem 4.13 of \cite{Choquet-Bruhat:GR}).
Hence we have
\begin{equation}
\label{eq:KomarMass}
M_{\rm ADM}=\frac{c^2}{8\pi G}\int_{S^2_\infty}\star dK^\flat\,.
\end{equation}
Here $K=\partial/\partial(ct)$, and 
$K^\flat:=g(K,\cdot)=-\Psi^2 cdt$ is the corresponding 1-form.
The star, $\star$, denotes the Hodge-duality map. Using 
(\ref{eq:PhiPsiRelation}) and asymptotic flatness it is 
now straightforward to show that the right hand side of 
(\ref{eq:KomarMass}) can indeed be written in the form of the 
middle term in (\ref{eq:DefGravMassNew}). This term only 
depends on $\Phi$ at infinity, i.e. the Newtonian limit,
and hence gives a value independent of the factor-2 
discrepancy discussed above. In this sense we may say that 
the active gravitational mass $M_g$ defined earlier 
corresponds to $M_{\rm ADM}$ in the GR context.

This ends our discussion of Einstein's 1912 scalar 
field equation, which is thus seen to contain many 
interesting features we know from GR, albeit in a 
pseudo Newtonian setting.

\bigskip
\noindent
\textbf{Acknowledgements.}
I sincerely thank the organisers and in particular 
Ji{\v r}\'{\i} Bi{\v c}{\'a}k
for inviting me to the most stimulating and beautiful conference 
\emph{Relativity and Gravitation 100 years after Einstein in Prague.} 

\end{document}